\documentclass[12pt]{article}
\usepackage{epsfig,amssymb,amsmath,psfrag}

\catcode`\@=11
\def \be  {\begin{equation}}
\def \ee  {\end{equation}}
\def \ba  {\begin{eqnarray}}
\def \ea  {\end{eqnarray}}
\def \baa {\begin{eqnarray*}}
\def \eaa {\end{eqnarray*}}
\def \bb  {\begin {thebibliography} }
\def \eb  {\end{thebibliography}}
\def \lab #1 {\label{#1}}

\newcommand\re[1]{(\ref{#1})}

\def \matrix #1 {\left(\begin{array}{cc} #1 \end{array}\right)}

\newcommand\lr[1]{{\left({#1}\right)}}

\newcommand \vev [1] {\langle{#1}\rangle}

\newcommand{\as}{\ifmmode\alpha_{\rm s}\else{$\alpha_{\rm s}$}\fi}
\newcommand{\asbar}{\ifmmode\bar{\alpha}_{\rm s}\else{$\bar{\alpha}_{\rm s}$}\fi}

\font\cmss=cmss12 
\def\inbar{\,\vrule height1.5ex width.4pt depth0pt}
\def\IC{\relax\hbox{$\inbar\kern-.3em{\rm C}$}}
\def\IZ{\relax{\hbox{\cmss Z\kern-.4em Z}}}
\def\IR{{\hbox{{\rm I}\kern-.2em\hbox{\rm R}}}}

\def\IP{{\hbox{{\rm I}\kern-.2em\hbox{\rm P}}}}
\def\II{\hbox{{1}\kern-.25em\hbox{l}}}

\begin{document}

\thispagestyle{empty}
\null\vskip-12pt \hfill  LAPTH-1230/07 \\
\null\vskip-12pt \hfill LPT--Orsay--07--136
\vskip2.2truecm
\begin{center}
\vskip 0.2truecm {\Large\bf
{\Large The hexagon  Wilson loop and \\[2mm] the BDS ansatz for the six-gluon
amplitude}
}\\
\vskip 1truecm
{\bf J.M. Drummond$^{*}$, J. Henn$^{*}$, G.P. Korchemsky$^{**}$ and E. Sokatchev$^{*}$ \\
}

\vskip 0.4truecm
$^{*}$ {\it Laboratoire d'Annecy-le-Vieux de Physique Th\'{e}orique
LAPTH\footnote{UMR 5108 associ\'{e}e \`{a}
 l'Universit\'{e} de Savoie},\\
B.P. 110,  F-74941 Annecy-le-Vieux, France\\
\vskip .2truecm $^{**}$ {\it
Laboratoire de Physique Th\'eorique%
\footnote{Unit\'e Mixte de Recherche du CNRS (UMR 8627)},
Universit\'e de Paris XI, \\
F-91405 Orsay Cedex, France
                       }
  } \\
\end{center}

\vskip 1truecm 
\centerline{\bf Abstract} 

\medskip

 \noindent
As a test of the gluon scattering amplitude/Wilson loop duality, we evaluate the
hexagonal light-like Wilson loop at two loops in $\mathcal{N}=4$ super Yang-Mills
theory. We compare its finite part to the Bern-Dixon-Smirnov (BDS) conjecture for
the finite part of the six-gluon amplitude. We find that the two expressions have
the same behavior in the collinear limit, but they differ by a non-trivial
function of the three (dual) conformally invariant variables. This implies that
either the BDS conjecture or the gluon amplitude/Wilson loop duality fails for
the six-gluon amplitude, starting from two loops. Our results are in
qualitative agreement with the
analysis of Alday and Maldacena of scattering amplitudes with infinitely
many external gluons.

\newpage
\setcounter{page}{1}\setcounter{footnote}{0}

\section{Planar gluon amplitude/Wilson loop duality}

With recent advances of the AdS/CFT correspondence, it became possible
to study gluon scattering amplitudes in maximally supersymmetric
Yang-Mills theory (SYM) both at weak and strong coupling.

At weak coupling, the conjecture was put forward by Bern, Dixon and Smirnov
\cite{bds05} that the maximally helicity-violating (MHV) amplitudes in
$\mathcal{N}=4$ SYM have a remarkable all-loop iterative structure. The
color-ordered planar $n-$gluon amplitude, divided by the tree amplitude, takes
the following form,
\be\label{BDS}
\ln\mathcal{M}_n = \text{[IR divergences]} + F_n^{\rm
  (MHV)}(p_1,\ldots,p_n) + O(\epsilon)\ .
\ee
Here the first term on the right-hand side describes infrared divergences in the
dimensional regularization scheme with $D=4-2\epsilon$, while the second term is
the finite contribution (dependent on the gluon momenta and the 't Hooft coupling
$a= {g^2 N}/(8\pi^2)$). The BDS conjecture provides an explicit expression for
the finite part, $F_n^{\rm (MHV)}=F_n^{\rm (BDS)}$, for an arbitrary number $n$
of external gluons, to all orders in the 't Hooft coupling. At present, the BDS
conjecture has been confirmed up to three loops for $F_4$ \cite{bds05} and up to
two loops for $F_5$ \cite{5point}. An explicit verification of the conjecture for
$n=6$ at two loops has not been performed up to now. However, we can at least say
that were the BDS conjecture \re{BDS} to fail, it would have to be corrected by
terms that satisfy an important additional consistency requirement. It originates
from the known two-loop asymptotic behavior of the scattering amplitude in the
collinear limit when the momenta of two neighboring on-shell gluons become
collinear. In this limit, $\mathcal{M}_n$ factorizes into the product of the
$(n-1)$-gluon amplitude and the universal splitting
amplitude~\cite{Bern:1994zx,Anastasiou:2003kj}. Since the BDS conjecture does
have this property \cite{bds05}, any potential correction to $F_n^{\rm (BDS)}$
must vanish in this limit.

Recently, Alday and Maldacena proposed the strong coupling description of
$n-$gluon scattering amplitudes \cite{am07} using the AdS/CFT correspondence.
According to their proposal, $\ln \mathcal{M}_n$ is given by the minimal surface
in AdS${}_5$ attached to a contour $C_n$, made of $n$ light-like segments
$[x_i,x_{i+1}]$, with the coordinates $x_i$ related to the on-shell gluon
momenta, $x_i^\mu-x_{i+1}^\mu=p_i^\mu$,
\be\label{AM}
\ln \mathcal{M}_n = -\frac{\sqrt{a}}{2\pi} A_{\rm min}(C_n)\,.
\ee
Remarkably, for $n=4$ the corresponding minimal surface can be found explicitly
and, after regularization, it leads to the same expression for $\ln\mathcal{M}_4$ as
eq.~\re{BDS} with the finite part $F_4$ in agreement with the BDS ansatz. For $n\ge
5$ the practical evaluation of the solution of the classical string
equations turns out to be difficult,  but it simplifies significantly for $n$ very
large \cite{Alday:2007he}. In the limit $n\to\infty$ the strong coupling prediction for $\ln\mathcal{M}_n$
 disagrees with the BDS conjecture \cite{Alday:2007he}.

Alday and Maldacena pointed out \cite{am07} that their prescription \re{AM} is
mathematically equivalent to the calculation of a Wilson loop at strong coupling \cite{M98,Kr02}.
Inspired by this, in \cite{DKS07} three of us conjectured that
a duality relation between planar gluon amplitudes and light-like Wilson loops also exists at weak
coupling. We illustrated this relation by an explicit one-loop calculation for
$n=4$. This was later extended to the case of arbitrary $n$ at one loop in
\cite{BHT07}. The duality relation reads
\be
\ln \mathcal{M}_n = \ln W(C_n) + O(\epsilon)\,, \label{duality}
\ee
with $C_n$ the same contour as before. We have recently verified this duality
at two loops for $n=4$ and $n=5$ and derived a conformal Ward identity for
the light-like Wilson loop $W(C_n)$, valid to all orders in the coupling
\cite{Drummond:2007cf,Drummond:2007au}. This Ward identity fixes the form of the
finite part of the Wilson loop for $n=4$ and $n=5$, up to an additive constant,
to agree with the conjectured BDS form for the corresponding gluon amplitudes. However, for $n\ge 6$ it allows for an arbitrary function of the conformal
invariants in addition to the BDS form (for $n=6$ there are three such
invariants). It is the purpose of the present letter to determine this function for $n=6$ at two loops.

The basic object we consider is the Wilson loop in
$\mathcal{N}=4$ SYM,
\begin{equation}\label{W}
    W\lr{C_n} = \frac1{N}\vev{0|\,{\rm Tr}\, \textrm{P}
    \exp\lr{i\oint_{C_n} dx^\mu A_\mu(x)}
    |0}\,,
\end{equation}
where $ A_\mu(x)=A_\mu^a(x) t^a$ is a gauge field and $t^a$ are the generators of the gauge group
$SU(N)$ in the fundamental representation. We use the conventions of \cite{Drummond:2007cf,Drummond:2007au} and
refer the interested reader to these papers for details. Even though
$\mathcal{N}=4$ SYM is a finite gauge theory, the Wilson loop \re{W} has specific
ultraviolet divergences due to the presence of cusps on the integration contour
$C_n$~\cite{P80,Brandt:1981kf,KR87}. To regularize these  singularities we use
dimensional regularization with $D=4-2\epsilon$. Like the scattering
amplitude, the Wilson loop can be factorized into divergent and finite parts,
\begin{equation}\label{W=ZF}
\ln W(C_n) = Z_n + F_n^{\rm (WL)} .
\end{equation}
Due to exponentiation of the cusp singularities  to all loops, the divergent part
$Z_n$ has the special form~\cite{KK92}
\begin{equation}\label{5}
   Z_n = -\frac{1}{4}  \sum_{l\ge 1}
   a^l\sum_{i=1}^n\lr{-x_{i-1,i+1}^2\mu^2}^{l\epsilon}
   \left[{\frac{\Gamma_{\rm
cusp}^{(l)}}{(l\epsilon)^2}+ \frac{\Gamma^{(l)}}{l\epsilon}}\right] \,,
\end{equation}
where $\Gamma_{\rm cusp}^{(l)}$ and $\Gamma^{(l)}$ are the expansion
coefficients of the cusp anomalous dimension and the so-called
collinear anomalous dimension, respectively, defined in the {adjoint}
representation of $SU(N)$:
\begin{align}\label{cusp-2loop}
& \Gamma_{\rm cusp}(a)=\sum_{l\ge 1} a^l\, \Gamma_{\rm cusp}^{(l)} = 2a -
\frac{\pi^2}3 a^2 + O(a^3)\ ,
\\ \notag
& \Gamma(a) =\sum_{l\ge 1} a^l\, \Gamma^{(l)} = - 7\zeta_3 a^2 + O(a^3)\,.
\end{align}
{ In \cite{Drummond:2007cf,Drummond:2007au} we confirmed these relations by an explicit two-loop
calculation of the divergent part of $W_4$ and $W_5$.}

The duality relation \re{duality} implies that upon a specific identification of the
regularization parameters, the infrared divergences of the scattering amplitude
$\mathcal{M}_n$  match the ultraviolet divergences of the light-like Wilson
loop $W(C_n)$ and, most importantly, the finite parts of the two objects
also coincide up to an inessential additive constant,
\be\label{finiteduality}
F_n^{\rm (MHV)} = F_n^{\rm (WL)} +\text{const}\ .
\ee
While the former property
immediately follows from the known structure of divergences of scattering
amplitudes/Wilson loops in a generic gauge theory \cite{KR87}, the latter
property \re{finiteduality} is extremely non-trivial.

In this letter we report on the two-loop calculation of $F_6^{\rm (WL)}$. We find that $F_6^{\rm (WL)} \neq
F_6^{\rm (BDS)}$,  with their difference being a non-trivial conformally invariant
function of the gluon momenta. At the same time, $F_6^{\rm (WL)}$ has the
same  collinear limit behavior as the six-gluon amplitude
$F_6^{\rm (MHV)}$.

\section{Finite part of the hexagon Wilson loop}

The finite part of the hexagon Wilson loop, $F_6^{\rm (WL)}$, does not depend on
the renormalization scale and it is a dimensionless function of the distances
$x_{ij}^2$. Since the edges of $C_6$ are light-like, $x_{i,i+1}^2=0$, the only
nonzero distances are $x_{i,i+2}^2$ and $x_{i,i+3}^2$ (with $i=1,\ldots, 6$ and
the periodicity condition $x_{i+6}=x_i$). We argued in
\cite{Drummond:2007cf,Drummond:2007au} that the conformal symmetry of the Wilson loop in
$\mathcal{N}=4$ SYM imposes severe constraints on $F_n^{\rm (WL)}$. It has to
satisfy the following Ward identity,
\begin{equation}\label{SCWI}
    \sum^n_{i=1} (2x_i^\nu x_i\cdot\partial_i - x_i^2 \partial_i^\nu)
 {F}_{n}  =
 \frac12\Gamma_{\rm cusp}(a) \sum_{i=1}^n  \ln
 \frac{x_{i,i+2}^2}{x_{i-1,i+1}^2} x^\nu_{i,i+1}\ .
\end{equation}
Specified to $n=6$, its general solution is given by \cite{Drummond:2007cf}
\be\label{solutionWI}
F_6^{\rm (WL)} = F_6^{\rm (BDS)} + f(u_1,u_2,u_3)\ .
\ee
Here, upon the identification $p_i=x_i-x_{i+1}$,
\begin{align}\label{BDS6point}
  F_{6} ^{\rm (BDS)}   =  \frac{1}{4} \Gamma_{\rm
 cusp}(a)\sum_{i=1}^{6} &\bigg[
    - \ln\Bigl(
\frac{x_{i,i+2}^2}{x_{i,i+3}^2} \Bigr)\ln\Bigl(
\frac{x_{i+1,i+3}^2}{x_{i,i+3}^2} \Bigr)
\nonumber\\
 &  +\frac{1}{4} \ln^2 \Bigl( \frac{x_{i,i+3}^2}{x_{i+1,i+4}^2}
\Bigr)   -\frac{1}{2} {\rm{Li}}_{2}\Bigl(1-
 \frac{x_{i,i+2}^2  x_{i+3,i+5}^2}{x_{i,i+3}^2 x_{i+2,i+5}^2}
\Bigr) \bigg]\,,
\end{align}
while $f(u_1,u_2,u_3)$ is an arbitrary function of the three
cross-ratios \footnote{The last term in \re{BDS6point} is a function
of cross-ratios only, but we keep it in $F_{6} ^{\rm (BDS)}$, because it is part of
the BDS conjecture.}
\begin{equation}u_1 = \frac{x_{13}^2 x_{46}^2}{x_{14}^2
x_{36}^2}, \qquad u_2 = \frac{x_{24}^2 x_{15}^2}{x_{25}^2 x_{14}^2},
\qquad u_3 = \frac{x_{35}^2 x_{26}^2}{x_{36}^2 x_{25}^2}\ . \label{u1u2u3}
\end{equation}
These variables are invariant under conformal transformations of the coordinates
$x_i^\mu$ and, therefore, they are annihilated by the conformal boost operator
entering the left-hand side of \re{SCWI}. In addition, the Wilson loop $W(C_6)$
is invariant under cyclic ($x_i \to x_{i+1}$) and mirror ($x_i\to x_{6-i}$)
permutations of the cusp points~\cite{Drummond:2007au}. This implies that
$f(u_1,u_2,u_3)$ is a totally symmetric function of three variables.

Combining together \re{finiteduality} and \re{solutionWI}, we conclude that were
the BDS conjecture {\it and}  the duality relation \re{finiteduality} correct for
$n=6$, we would expect that $f(u_1,u_2,u_3)=\text{const}$. The explicit two loop
calculation we report on here shows that this is not true.

For the sake of simplicity we performed the calculation of $W(C_6)$ in the Feynman
gauge. In addition, we made  use of the non-Abelian exponentiation
property of Wilson loops \cite{Gatheral} to reduce the number of
relevant Feynman diagrams. In application to  $f(u_1,u_2,u_3)$ this property can
be formulated as follows (the same property also holds for $F_6^{\rm (WL)}$)
\begin{equation}\label{W-decomposition}
f = \frac{g^2}{4\pi^2}C_F\,
f^{(1)}  +  \lr{\frac{g^2}{4\pi^2}}^2 C_F N\, f^{(2)}  +
O(g^6)\,,
\end{equation}
where $C_F=t^a t^a=(N^2-1)/(2N)$ is the Casimir in the fundamental
representation of the $SU(N)$. The functions $f^{(1)}$ and $f^{(2)}$
do not involve the color factors and only depend on the distances
 between the cusp points on $C_6$.
At one loop, $f^{(1)}(u_1,u_2,u_3)$ is in fact a constant \cite{BHT07}.

\begin{figure}[h]
\psfrag{1}{}\psfrag{2}{}\psfrag{3}{}\psfrag{4}{}\psfrag{5}{}\psfrag{6}{}\psfrag{7}{}\psfrag{8}{}
\psfrag{9}{}\psfrag{10}{}\psfrag{11}{}
\psfrag{12}{}\psfrag{13}{}\psfrag{14}{}\psfrag{15}{}\psfrag{16}{}\psfrag{17}{}\psfrag{18}{}
\psfrag{19}{}\psfrag{20}{}\psfrag{21}{}
\psfrag{x1}[rc][cc]{$\scriptscriptstyle x_1$}
\psfrag{x2}[rc][cc]{$\scriptscriptstyle x_6$}
\psfrag{x3}[rc][cc]{$\scriptscriptstyle x_5$}
\psfrag{x4}[lc][cc]{$\scriptscriptstyle x_4$}
\psfrag{x5}[lc][cc]{$\scriptscriptstyle x_3$}
\psfrag{x6}[lc][cc]{$\scriptscriptstyle x_2$}
\centerline{{\epsfxsize17cm \epsfbox{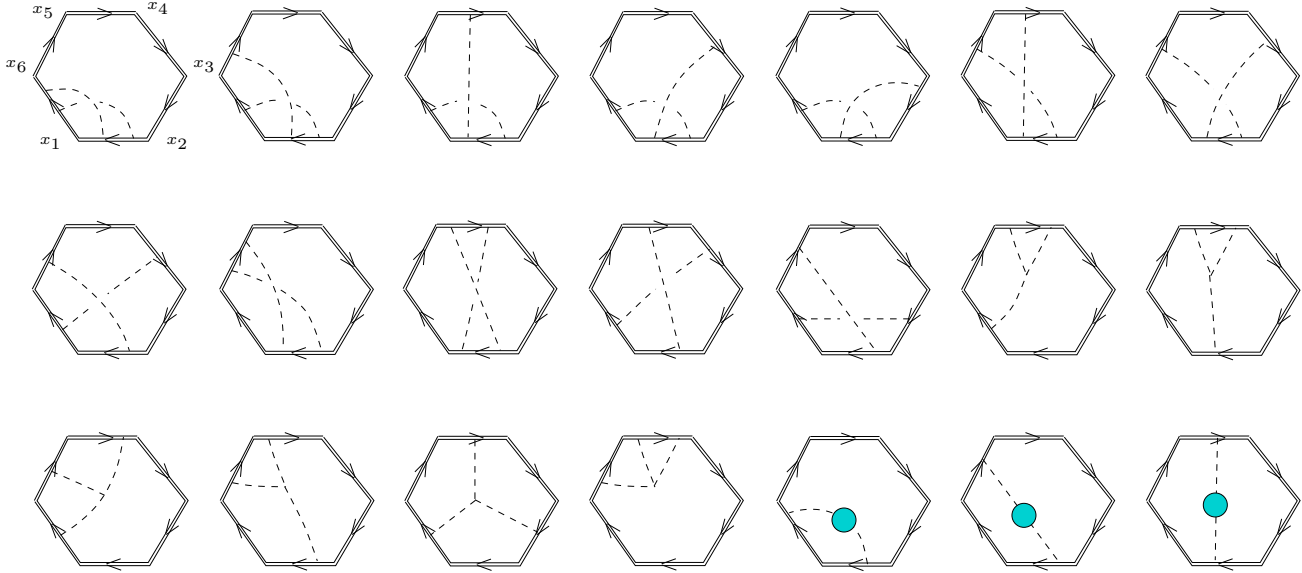}}} \caption[]{\small The
maximally non-Abelian Feynman diagrams of different topology contributing to
$F_6^{\rm (WL)}$. The double lines depict the integration contour $C_6$, the
dashed lines -- the gluon propagator and the blob -- the one-loop polarization operator.
}
\label{Fig-WL6}
\end{figure}

\begin{figure}[h]%
\psfrag{gamma}[cc][cc]{$\gamma$}
\psfrag{f}[lc][cc]{}
\centerline{{\epsfxsize9cm \epsfbox{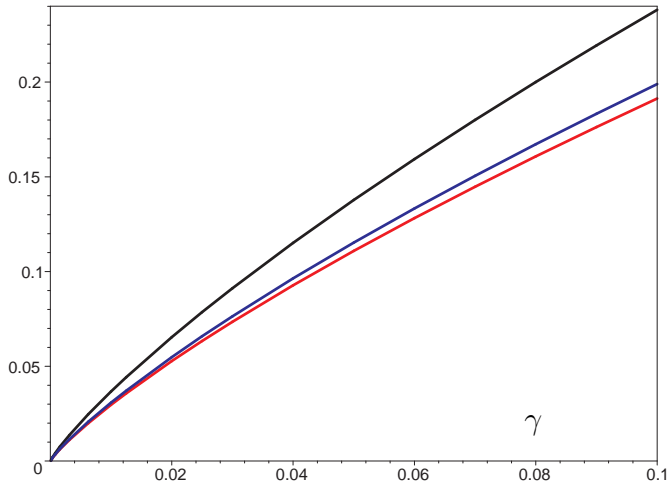}}} \caption[]{\small The
$\gamma-$dependence of the function $\widehat f^{(2)}(\gamma,u,1-u)$,
Eq.~\re{sub}, for
  different values of
the parameter $u=0.5$ (lower curve), $u=0.3$ (middle curve) and $u=0.1$ (upper
curve).}
\label{Fig-f}
\end{figure}

\begin{figure}[h]%
\psfrag{u}[cc][cc]{\vspace*{-10mm}$u$}
\psfrag{f}[lc][cc]{ }
\centerline{{\epsfxsize9cm \epsfbox{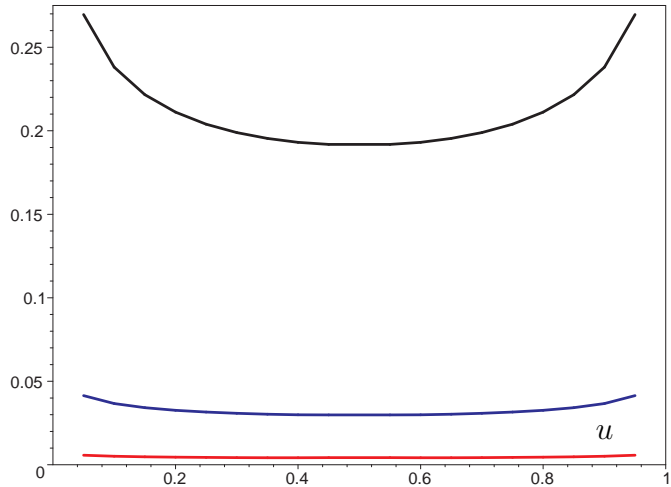}}} \caption[]{\small The
$u-$dependence of the function $\widehat f^{(2)}(\gamma,u,1-u)$, Eq.~\re{sub},
for different values of the parameter $\gamma=0.001$ (lower curve), $\gamma=0.01$
(middle curve) and $\gamma=0.1$ (upper curve).}
\label{Fig-f2}
\end{figure}

As explained in \cite{Drummond:2007cf}, the relation \re{W-decomposition} implies
that in order to determine the function $F_6^{\rm (WL)}$ at two loops (and hence $f^{(2)}$) it is sufficient to calculate
the contribution to $W(C_6)$ from two-loop diagrams containing the `maximally
non-Abelian' color factor $C_F N$ only. All relevant two-loop graphs are shown in
Fig.~\ref{Fig-WL6}.
We derived parameter integral representations for all the Feynman graphs. The
integrals are difficult to evaluate analytically and so we calculated them
numerically for many different sets of values of $x_{ij}^2$.\footnote{One should bear in mind that the allowed values of $x_{ij}^2$ have to obey kinematical constraints. They originate from the six gluon momenta $p^\mu_i$ satisfying  $p^2_i=0$ and $\sum_{i=1}^6 p^\mu_i = 0$. Solving these constraints is not a trivial task. We are grateful to Fernando Alday for sharing with us his numerical solutions for the kinematical configurations.} We found that, firstly, for
values of $x_{ij}^2$ related by conformal boosts (hence leaving
$u_{1},u_{2},u_{3}$ invariant),  the difference $F_6^{\rm (WL)}- F_6^{\rm (BDS)}$
remains constant. Thus, it only depends on the cross-ratios \re{u1u2u3}, in
agreement with \re{solutionWI}. Secondly, varying the values of the cross-ratios
we found that $f^{(2)}(u_{1},u_{2},u_{3}) \neq \text{const}$ (see
Fig.~\ref{Fig-f} and Fig.~\ref{Fig-f2}), i.e. it is a non-trivial function of $u_{1},u_{2},u_{3}$.

This means that either the BDS conjecture, or the gluon
amplitude/Wilson loop duality (or both) is not correct for $n=6$,
starting from two loops.  At this stage, we cannot discriminate
between the different scenarios. Nevertheless, we can show that
$F_6^{\rm (WL)}$ has the same collinear limit behavior as $F_6^{\rm
  (BDS)}$ at two loops, i.e. $f^{(2)}(u_{1},u_{2},u_{3})$ tends to a
constant in the collinear limit.

We recall that for the six-gluon amplitude $\mathcal{M}_6$ depending on
light-like momenta, $\sum_{i=1}^6 p_i^\mu=0$ and $p_i^2=0$, the collinear limit
amounts to letting, e.g. $p^{\mu}_{5}$ and $p^{\mu}_{6}$ be nearly
collinear (see e.g. \cite{Bern:1994zx} for more details), so that
$(p_5+p_6)^2 \to 0$ and
\be
p_5^\mu \to z P^\mu\,,\qquad p_6^\mu \to (1-z)  P^\mu\,,
\ee
with $P^2=0$ and $0< z < 1$ being the momentum fraction. Using the identification
$p_i^\mu = x_i^\mu -x_{i+1}^\mu$, we translate these relations into
properties of
the corresponding Wilson loop $W(C_6)$. We find that the cusp at point
$6$ is `flattened' in the collinear limit and the contour $C_6$
reduces to one with five cusps. In terms of the distances
$x_{ij}^2$, the collinear limit amounts to
\begin{align}\label{collinear1}
x_{15}^2 & \rightarrow   0 \,, & & x_{36}^2 \rightarrow z x_{13}^2 + (1-z)
x_{35}^2 \nonumber  \,, \\[2mm]
 x_{46}^2 &\rightarrow  z
x_{14}^2\,, & & x_{26}^2 \rightarrow (1-z) x_{25}^2 \,,
\end{align}
while the other distances $x_{13}^2,x_{24}^2,x_{25}^2,x_{35}^2$ remain
unchanged.
For the conformal cross-ratios the relation \re{collinear1} implies
\begin{equation}\label{collinear2}
u_{1} \rightarrow u \,,\quad u_{2} \rightarrow 0 \,,\quad u_{3} \rightarrow
1-u\,,
\end{equation}
with $u= {z x_{13}^2}/{(z x_{13}^2 +(1-z) x_{35}^2)}$ fixed. As was already
mentioned, the relation \re{solutionWI} is consistent with the
collinear limit of
the six-gluon amplitude provided that, in the limit \re{collinear2},
the function
$f(u_1,u_2,u_3)$ approaches a finite value independent of the kinematical
invariants. The same property can be expressed as follows (we recall that the function $f(u_1,u_2,u_3)$ is totally symmetric)
\be\label{const}
f(0,u,1-u) = c \,,
\ee
with $c$ being a constant.
Using our two-loop results for the finite part $F_6$, we performed thorough
numerical tests of the relation \re{const} for different kinematical
configurations of the contour $C_6$.

We found that, in agreement with \re{const}, the limiting value of the
function  $f^{(2)}(\gamma,u,1-u)$ as $\gamma\to 0$ does not depend on $u$. Since
the duality relation \re{finiteduality} is not sensitive to the value of this constant, it is
convenient to subtract it from $f^{(2)}(\gamma,u,1-u)$ and introduce the function
\be\label{sub}
\widehat f^{(2)}(\gamma,u,1-u) = c - f^{(2)}(\gamma,u,1-u)
\ee
which satisfies  $\widehat f^{(2)}(0,u,1-u)=0$. To summarize our findings,  in Fig.~\ref{Fig-f} we
plot the function $\widehat f^{(2)}(\gamma,u,1-u)$ against
$\gamma$ for different choices of the parameter $0<u<1$ and in Fig.~\ref{Fig-f2}
the same function against $u$ for different choices of the parameter $\gamma$.
The important region for the collinear limit is where $\gamma$ is close to zero.
We also give numerical values for a  range of values of $\gamma$ such that
one can see how the function $f^{(2)}(u_1 ,u_2 , u_3 )$ varies in the particular
parametrization $u_{1} = \gamma, u_{2} =u , u_{3}=1-u$.

\section{Conclusions}

Given the results we have presented in this letter, it is urgent to know the
six-gluon amplitude at two loops. Depending on the outcome of this calculation, we can envisage the following three scenarios:

\begin{itemize}

    \item If the duality between amplitudes and Wilson loops persists for the
six-gluon amplitude at two loops, then the BDS conjecture fails and the
difference will be given by the function $f^{(2)}(u_1,u_2,u_3)$ that we
have found.

    \item If the BDS conjecture holds, then the duality between amplitudes
and Wilson loops breaks down for $n=6$ at two loops.

    \item  If the gluon amplitude disagrees with both the BDS ansatz and the Wilson
loop, then it would be very interesting to verify whether it still respects dual
conformal symmetry \cite{Drummond:2006rz,Drummond:2007cf,Drummond:2007au} (i.e. the difference is a function of the conformal
cross-ratios).

\end{itemize}

The finite part of the one-loop MHV amplitude involves functions of the kinematical invariants of transcendentality $2$ (double logs and dilogs).
We expect that this is a general feature, i.e. the
finite part of $\ln \mathcal{M}_n$ should have maximal transcendentality $2\ell$ at $\ell$ loops. This is indeed true for the BDS ansatz \re{BDS6point}, where the non-trivial functions are of
transcendentality $2$ and the factor $\Gamma_{\rm
 cusp}(a)$ is supposed to supply the remaining transcendentality $2(\ell-1)$. There are \textit{a priori} no reasons why
functions of higher transcendentality should not appear at higher loops, provided that they have the general analyticity properties of gluon amplitudes, including the correct collinear limit behavior. An example for this is the constant term in the finite part of $\ln W(C_n)$, as we have  demonstrated by explicit two loop
calculations for $n=4$ and $n=5$. We conjecture that the same property holds for arbitrary $n$ to all
orders. In particular, we expect that our two-loop function $F_6^{\rm (WL)} - F_6^{\rm
(BDS)}=f(u_1,u_2,u_3)$ has
transcendentality $4$. Needless to say, it would be very interesting to identify
its analytical form.

Independently of the outcome of the two-loop calculation of the
six-gluon amplitude, we have presented an example of a non-trivial
function which is not captured by the BDS ansatz and which has
the right collinear limit properties to appear in the final
two-loop expression for the six-gluon amplitude.

\section*{Acknowledgements}

We would like to thank  F.~Alday, A.~Belitsky, Z.~Bern, L.~Dixon, A.~Gorsky, J.~Maldacena, J.~Plefka and
V.~Smirnov for stimulating discussions.
This research was supported in part by the French Agence
Nationale de la Recherche under grant ANR-06-BLAN-0142.


 \end{document}